\documentstyle[epsfig]{mn}

\input{psfig.sty}

\ifnfsstwo

\fi

\ifnfssone
  \newmathalphabet{\mathit}
    \addtoversion{normal}{\mathit}{cmr}{m}{it}
    \addtoversion{bold}{\mathit}{cmr}{bx}{it}

\fi

\ifoldfss

\fi

\loadboldmathitalic
\loadboldgreek

\title[Evolution of Ba and Eu in dSph]
       {The Evolution of Barium and Europium in Local Dwarf Spheroidal
Galaxies}
\author[G. Lanfranchi, F. Matteucci and G. Cescutti]
 {Gustavo A. Lanfranchi $^1$,  Francesca Matteucci$^2$ and Gabriele 
Cescutti$^2$\\
$^1$IAG-USP,
 R. do Mat\~ao 1226, Cidade Universit\'aria, 05508-900 S\~ao Paulo, 
SP, Brazil\\
 $^2$ Dipartimento di Astronomia-Universit\'a di Trieste,
  Via G. B. Tiepolo 11, 34131 Trieste, Italy}

\pubyear{2005}

\begin{document}
\maketitle

\begin{abstract}
By means of a detailed  chemical evolution model, we follow the evolution 
of barium and europium in four Local Group
Dwarf Spheroidal Galaxies,
in order to set constraints on the nucleosynthesis of these
elements and on the evolution of this type of galaxies compared 
with the Milky Way. The model, which is able to reproduce several 
observed abundance ratios and the present day total mass and gas 
mass content of these galaxies, adopts up to date nucleosynthesis 
and takes into account the role played by supernovae of different 
types (II, Ia) allowing us to follow in detail the evolution of 
several chemical elements (H, D, He, C, N, O, Mg, Si, S, Ca, 
Fe, Ba and Eu). By assuming that barium is a neutron capture 
element produced in low mass AGB stars by s-process 
but also in massive stars (in the mass range 10 - 30 $M_{\odot}$) 
by r-process,
during the explosive event of supernovae of type II, and that 
europium is a pure r-process element synthesized in massive stars 
also in the range of masses 10 - 30 $M_{\odot}$, we are able to 
reproduce the observed [Ba/Fe] and [Eu/Fe] as functions of [Fe/H] 
in all four  galaxies studied. We confirm also the important role 
played by the very low star formation efficiencies ($\nu$ = 0.005 
- 0.5 Gyr$^{-1}$) and by the intense galactic winds (6-13 times the 
star formation rate) in the evolution of these galaxies. These low 
star formation 
efficiencies (compared to the one for the Milky Way disc) adopted 
for the Dwarf Spheroidal Galaxies are the main reason 
for the differences between the trends of [Ba/Fe] and [Eu/Fe] 
predicted  and observed in these galaxies and in the metal-poor 
stars of our 
Galaxy. Finally, we provide predictions for Sagittarius galaxy 
for which data of only two stars are available.

\end{abstract}

\begin{keywords}
stars: abundance -- stars: heavy elements -- galaxies: abundance 
ratios -- galaxies: Local Group -- galaxies: evolution -- 
\end{keywords}

\section{Introduction}

The proximity and the relative simplicity of the Local Group (LG) 
Dwarf Spheroidal (dSph) Galaxies make these systems excellent 
laboratories to test assumptions regarding the nucleosynthesis of
chemical elements and theories of galaxy evolution. Several 
studies addressing the observation of red giants stars in local 
dSph galaxies with high resolution spectroscopy allow one to 
infer accurately the abundances of several elements including 
$\alpha$-, iron-peak and very heavy elements, 
such as barium and europium (Smecker-Hane $\&$ McWilliam 
1999; Bonifacio et al. 2000; Shetrone, Cot\'e $\&$ Sargent 
2001; Shetrone et al. 2003; Bonifacio et al. 2004; Sadakane 
et al. 2004; Geisler et al. 2005). These abundances and abundance 
ratios are not only central ingredients in galactic chemical 
evolution studies but are also very important in the attempt 
to clarify some aspects of the processes responsible for the
formation of chemical elements.

There is a certain consensus regarding the general aspects of the
production of iron peak and $\alpha$- elements, but for some other 
elements several questions concerning their production are still
open for debate (Fran\c cois et al. 2004). It is largely 
accepted that $\alpha$- elements are mainly
produced on short time-scales in explosions of type II 
supernovae (SNe II, Woosley $\&$ Weaver 1995; Thielemann, Nomoto $\&$
Hashimoto 1996; Nomoto et al. 1997), whereas the iron-peak ones are 
the main products of type Ia supernovae (SNe Ia, Nomoto et al. 1997), 
which occur on longer time-scales. The different time-scales 
involved in these explosions are the main reason for using the 
ratio between these two sets of elements (the [$\alpha$/Fe] ratio)
as a cosmic clock, which can be used to impose constraints in 
the star formation (SF) history of the 
systems where this ratio is observed (Tinsley 1980; Matteucci 1996). 
On the other hand, the sites of formation and 
the processes which are responsible for the production of
elements heavier than Fe remain unclear. These elements are
synthesized mainly by neutron capture reactions and are normally
divided into r-process and s-process elements according to the 
velocity of the neutron capture reaction, rapid and slow being relative
to the duration of the $\beta$-decay process, respectively. 
Among these elements the production of Barium and Europium 
is more often discussed, due to the characteristics of their
formation. They are believed to be  produced by the two neutron 
capture processes in different ways: while Eu is claimed to 
be a pure r-process element it is believed that the process 
which dominates the Ba production is the s-process, but with 
a low fraction of Ba being produced also by the
r-process (Wheeler, Sneden $\&$ Truran 1989). The stellar mass 
ranges where these elements are formed are not yet fully understood. 
The r-process is generally accepted to take place in SNe II 
explosions (Hill et al. 2002; Cowan et al. 2002). 
The s-nuclei can be divided into the main component and 
the weak component elements.  The main component elements are synthesized
during the thermally pulsing asymptotic giant branch (AGB) phase of 
low mass stars (Gallino et al. 1998; Busso et al. 2001). A different 
site is required for the production of the weak 
component of the s-process elements. These elements are believed to 
be produced also in advanced evolutionary stages of massive stars 
(Raiteri et al. 1993).

Models of stellar evolution predict 
that the
main formation (by s-process) of Ba occurs  during the
thermal pulses of AGB stars with initial masses between 1 to 3 
$M_{\odot}$ (Gallino et al. 1998; Busso et al. 2001), with a negligible 
contribution from intermediate mass stars of 
($ 3 < M < 8 M_{\odot}$).  As a consequence, the Ba enrichment 
of the interstellar medium (ISM)
due to the s-process is delayed 
relative to the Fe enrichment, which is produced also in massive stars,
and one should expect a low [Ba/Fe]
at low metallicities, which would later increase due to the injection 
of s-processed Ba by the low mass stars (LMS) as [Fe/H] also increases. 
It has also been 
suggested that a low fraction of Ba is produced by 
r-process in stars with masses at the low end of high 
mass stars ($ 8 < M < 10 M_{\odot}$) (Travaglio et al. 2001), in order 
to reproduce the [Ba/Fe] ratios observed at low metallicities in the 
Milky Way. This production would contribute with Ba at early 
stages of the evolution of the galaxy, prior to the peak of the iron 
enrichment by SNe Ia. 

Europium, on the other hand, is produced only by the r-process
which is believed to occur in massive stars ($M > 8 M_{\odot}$)
(Woosley et al. 1994). The details of the r-process nucleosynthesis 
remain, however, unclear and several scenarios have been 
proposed (Woosley et al. 1994; 
Freiburghaus et al. 1999; Wanajo et al. 2003). Assuming
this site of production for Eu, one should observe a relatively
high [Eu/Fe] at low metallicities, which would then decrease
due to the injection of Fe in the ISM by SNe Ia explosions.
However, there is a large spread in the observational data
concerning both Eu and Ba in the stars of our Galaxy 
(Ryan, Norris $\&$ Beers 1996; Sneden et al. 1998; Burris et al. 
2000), which prevents one from drawing  very firm conclusions
regarding the production of both Eu and Ba in the Milky Way.

In local dSph galaxies, red giant stars have been observed with
high resolution spectroscopy and the abundances of Eu and Ba, 
among others elements, have been derived (Shetrone, Cot\'e $\&$ 
Sargent 2001; Shetrone et al. 2003; Sadakane et al. 2004; 
Geisler et al. 2005).

Shetrone, Cot\'e $\&$ 
Sargent (2001) argued that Draco and Ursa Minor stars exhibit an 
abundance pattern consistent with one dominated by the r-process, i.e. 
[Ba/Eu] ranges from solar values at high metallicities to [Ba/Eu] 
$\sim$ -0.5 at [Fe/H] $\leq$ -1 dex. The pattern of [Ba/Fe] and [Eu/Fe] 
also resembles the one observed in the halo field stars according to
these authors. The same conclusion was reached by Shetrone et al. (2003), 
who analysed these abundance ratios in Sculptor, Fornax and Carina. 
Shetrone et al. (2003) claimed also that in Sculptor, Fornax and Leo
I the 
pattern of [Eu/Fe] is consistent with the production of Eu in SNe II.

On the other hand, Venn et al. (2004), pointed out that, despite the 
general similarity, the dSph stars span a larger range in [Ba/Fe] 
and [Eu/Fe] ratios at intermediate metallicities than 
the Galactic stars and, more important, that about half of the dSph
stars exhibit lower [Y/Eu] and 2/3 higher [Ba/Y] than the Galactic stars
at the same metallicity, thus suggesting a clear difference between the
chemical evolution of our Galaxy and the one of dSph galaxies. 
The [$\alpha$/Fe] ratios observed in dSphs also are different from the same 
ratios in the Milky Way showing in general lower [$\alpha$/Fe] ratios than 
the Galactic stars with the same [Fe/H] (Smecker-Hane $\&$ McWilliam 
1999; Bonifacio et al. 2000; Shetrone, Cot\'e $\&$ Sargent 
2001; Shetrone et al. 2003; Bonifacio et al. 2004; Sadakane et al. 2004;
Geisler et al. 2005).

These observations not only shed some light into
the chemical evolution history of these galaxies but allowed also
the construction of chemical evolution models aimed at reproducing
important observational constraints, such as the elemental abundance 
ratios, the present gas mass and total mass (Carraro et al. 2001; 
Carigi, Hernandez $\&$ Gilmore 2002; Ikuta $\&$ Arimoto 2002; 
Lanfranchi $\&$ Matteucci 2003 (LM03); Lanfranchi $\&$ Matteucci
2004 (LM04)). 
Among these models the one proposed by LM03 and LM04
for 6 local dSph galaxies (namely Draco, Carina, Sculptor, Sextan,
Ursa Minor
and Sagittarius) succeeded  in reproducing the observed [$\alpha$/Fe] 
ratios, the present
gas mass and final total mass by adopting a very low star formation rate,
$\nu$ $\sim$ 0.01 to 0.5 Gyr$^{-1}$ (with lower values for Draco 
and higher ones for Sagittarius) and a high wind efficiency 
(6-13 times the star formation rate).  
Besides that, LM04 predicted the stellar metallicity
distribution of these galaxies which were later on compared to 
observational data for Carina with a reasonably good 
agreement (Koch et al. 2004).
The success of LM03 and LM04 models in reproducing several 
observational constraints allows us to use them as tools to test the
theories about the sites of production and the processes responsible for
the synthesis of Ba and Eu in dSph galaxies. By adopting the
nucleosynthesis prescriptions for these elements which are able to
reproduce the most recent observed data for our Galaxy (Cescutti 
et al. 2005)
and comparing the predictions of the models with observational data,
it is possible to verify if the assumptions made regarding the 
nucleosynthesis of Ba and Eu can also fit the data of local dSph 
galaxies.

The paper is organized as follows: in Sect. 2 we present
the observational data concerning the dSph galaxies,
in Sect. 3 the adopted chemical evolution models,
the star formation and the nucleosynthesis prescriptions 
are described, in Sect. 4 the predictions of our models are 
compared to the observational data and the results discussed, 
and finally in Sect. 5 we draw some conclusions. All elemental
abundances are normalized to the solar values
([X/H] =  log(X/H) - log(X/H)$_{\odot}$)  measured by 
Grevesse $\&$ Sauval (1998).

\section{Data Sample}

Recently, red giant stars of dSph galaxies have been the subject 
of several works with the aim of determining with high-resolution
spectroscopy the abundance of several chemical elements
including heavy elements such as barium and europium
(Bonifacio et al. 2000; Shetrone, Cot\'e $\&$ Sargent 2001; 
Shetrone et al. 2003; Venn et al. 2004; Sadakane et al. 2004; 
Fulbright, Rich $\&$ Castro, 2004; Geisler et al. 2005). 
From these observations we gathered the data from the galaxies
that were analysed in LM03 
and LM04 and for which there are abundance determinations for 
both Ba and Eu. They are Carina, Draco, Sculptor, Ursa Minor and Sagittarius.
Despite of the relative small number of data points, it is 
possible to compare the observed abundance ratios 
with the model predictions. We choose to compare the observed
ratios [Ba/Fe], [Eu/Fe] and [Ba/Eu] with the ones predicted by 
the models, since these ratios can provide some clues not only
to the nucleosynthesis of Ba and Eu, but also to all s-process
and r-process elements. 

In order to properly compare different 
data from different authors with the predictions of the models we adopted 
the abundance values of Shetrone, Cot\'e $\&$ Sargent (2001), 
Shetrone et al. (2003), and  Sadakane et al. (2004) updated by
Venn et al. (2004). Venn et al. (2004) homogenized the atomica data 
for spectral
lines of Ba and Eu providing data with improved quality which allow a
consistent comparison between data from different sources. Otherwise,
the effect
of combining these different data would be seen as a larger spread 
in the abundances and possibly in the abundance ratios of 0.1 to 0.2 dex
(see Venn et al. 2004). In the case of Bonifacio et al. (2000) data, Eu 
is obtained using hyper-fine splitting (HFS) (see their Table 5), but Ba 
is not. The authors claimed that the Ba abundances obtained with HFS would 
exhibit no significant difference since the line observed (Ba II 6496.9) 
is a strong line which is not affected by this correction (Bonifacio 
private communication, see also Shetrone et al. 2003).

Some of the observed stars, however, exhibit anomalous values of
[Ba/H] or [Eu/H], and for this reason were excluded from the sample. 
Two stars in Ursa Minor, K and 199 (in Shetrone, Cot\'e $\&$ 
Sargent 2001), exhibit heavy-element abundance ratios enhanced
relative to those typical for other dSph stars: the Ursa Minor K star
has an abundance pattern dominated by the s-process and
was classified as a Carbon star while Ursa Minor 199 is 
dominated by r-process (see also Sadakane et al. 2004). In Sculptor,
there are also two stars with enhanced heavy-element abundance:
Sc982 (Geisler et al. 2005) and Sculptor H-400 (Shetrone et al. 2003).
While Shetrone et al. (2003) claimed that the r-process dominated 
abundance could be attributed to inhomogeneous mixing of the SNe II
yields, Geisler et al. (2005) classified Sc982 as a heavy element star
which could have been enriched by an other star, which is now dead.
Either way, all these stars do not exhibit an abundance pattern
characterized only by the nucleosynthesis process occurring inside
the star, but also one which was contaminated by external factors. 
The maintenance of these stars in the sample could lead to
an erroneous comparison with the model predictions and, as 
a consequence, to a misleading interpretation and to wrong conclusions
regarding the processes and the site of production of the heavy
elements analysed. Therefore, we excluded these stars from our sample, 
whereas all the other stars were considered and included in 
the comparisons with the models predictions.

\begin{table*}
\begin{center}\scriptsize  
\caption[]{Models for dSph galaxies. $M_{tot}^{initial}$ 
is the baryonic initial mass of the galaxy, $\nu$ is the star-formation 
efficiency, $w_i$ is the wind efficiency, and $n$, $t$ and $d$ 
are the number, time of occurrence and duration of the SF 
episodes, respectively.}
\begin{tabular}{lccccccc}  
\hline\hline\noalign{\smallskip}  
galaxy &$M_{tot}^{initial} (M_{\odot})$ &$\nu(Gyr^{-1})$ &$w_i$
&n &t($Gyr$) &d($Gyr$) &$IMF$\\    
\noalign{\smallskip}  
\hline
Sculptor &$5*10^{8}$ &0.05-0.5 &11-15 &1 &0 &7 &Salpeter\\
Draco  &$5*10^{8}$ &0.005-0.1 &6-10 &1 &6 &4 &Salpeter\\
Ursa Minor &$5*10^{8}$ &0.05-0.5 &8-12 &1 &0 &3 &Salpeter\\
Carina &$5*10^{8}$ &0.02-0.4 &7-11 &2 &6/10 &3/3 &Salpeter\\
Sagittarius &$5*10^{8}$ &1.0-5.0 &9-13 &1 &0 &13 &Salpeter\\
\hline\hline
\end{tabular}
\end{center}
\end{table*} 

\section{Models} 

We use in this work the same chemical evolution model for dSphs
galaxies as described in LM03 and LM04. The model is able to reproduce
the [$\alpha$/Fe] ratios, the present gas mass and the inferred total mass
of six dSph galaxies of the Local Group, namely Carina, Draco, Sculptor, 
Sextan, Sagittarius and Ursa Minor, and also the stellar metallicity 
distribution of Carina (Koch et al. 2004). The scenario representing 
these galaxies is characterized by one long episode (two episodes
in the case of Carina) of star formation (SF) with very low efficiencies
(except in the case of Sagittarius) - $\nu$ = 0.001 to 0.5 Gyr $^{-1}$ -
and by the occurrence of very intense galactic winds - $w_i$ = 6-13. 
The model allows one to follow in detail the evolution of the 
abundances of several chemical elements, starting from 
the matter reprocessed by the stars and restored into the 
ISM by stellar winds and type II and Ia supernova explosions.

The main features of the model are:

\begin{itemize}

\item
one zone with instantaneous and complete mixing of gas inside
this zone;

\item
no instantaneous recycling approximation, i.e. the stellar 
lifetimes are taken into account;

\item
the evolution of several chemical elements (H, D, He, C, N, O, 
Mg, Si, S, Ca, Fe, Ba and Eu) is followed in detail;

\end{itemize}

In the scenario adopted in the previous works, the dSph galaxies 
form through
a continuous and fast infall of pristine gas until a mass of
$\sim 10^8 M_{\odot}$ is accumulated.  One crucial 
feature in the evolution of these galaxies is the occurrence 
of galactic winds, which develop when the thermal 
energy of the gas equates its binding energy (Matteucci $\&$
Tornamb\'e 1987). This quantity is strongly influenced by 
assumptions concerning the presence and distribution 
of dark matter (Matteucci 1992). A diffuse ($R_e/R_d$=0.1, 
where $R_e$ is the effective radius of the galaxy and $R_d$ is 
the radius of the dark matter core) but massive 
($M_{dark}/M_{Lum}=10$) dark halo has been assumed for each galaxy.

\subsection{Theoretical prescriptions} 

The evolution in time of the fractional mass of the element $i$ 
in the gas within a galaxy, $G_{i}$, is described by the basic 
equation:

\begin{equation}
\dot{G_{i}}=-\psi(t)X_{i}(t) + R_{i}(t) + (\dot{G_{i}})_{inf} -
(\dot{G_{i}})_{out}
\end{equation}

where $G_{i}(t)=M_{g}(t)X_{i}(t)/M_{tot}$ is the gas mass in 
the form of an element $i$ normalized to a total fixed mass 
$M_{tot}$ and $G(t)= M_{g}(t)/M_{tot}$ is the total fractional 
mass of gas present in the galaxy at the time t. The quantity 
$X_{i}(t)=G_{i}(t)/G(t)$ represents the abundance by mass 
of an element $i$, with the summation over all elements 
in the gas mixture being equal to unity. The star formation rate (SFR),
i.e. the fractional amount of gas turning into stars per unit 
time, is given by $\psi(t)$, while the returned fraction 
of matter in the form of an element $i$ that the stars 
eject into the ISM through stellar winds and supernova 
explosions is represented by $R_{i}(t)$. This term contains 
all the prescriptions 
concerning the stellar yields and the supernova progenitor 
models. The infall of external gas and
the galactic winds are accounted for by the two terms 
$(\dot{G_{i}})_{inf}$ and $(\dot{G_{i}})_{out}$, respectively.
The prescription adopted for the star formation history is the
main feature which characterizes the dSph galaxy models.

The SFR $\psi(t)$ has a simple form and is given by:

\begin{equation}
\psi(t) = \nu G(t)
\end{equation}
where $\nu$ is the inverse of the typical time-scale 
for star formation, the SF efficiency, and is expressed 
in $Gyr^{-1}$.

The star formation is not halted even after the onset of the 
galactic wind but proceeds at a lower rate since a large
fraction of the gas ($\sim\, 10\%$) is carried out of the galaxy. 
The details of the star formation, such as number of episodes, 
time of occurrence and duration, are taken from the star 
formation history of each individual galaxy as inferred by 
CMDs taken from Dolphin (2002) and Hernandez, Gilmore 
$\&$ Valls-Gabaud (2000). It is generally adopted 1 episode of
SF (2 in the case of Carina), with durations which
vary from 3 Gyr to 7 Gyr (see Table 1 for more details).
 
The rate of gas infall is defined as:
\begin{eqnarray}
(\dot G_{i})_{inf}\,=\,Ae^{-t/ \tau}
\end{eqnarray}
with A being a suitable constant and $\tau$ the infall 
time-scale which is assumed to be 0.5 Gyr.

The rate of gas loss via galactic winds for each element 
{\it i} is assumed to be proportional to the star formation 
rate at the time {\it t}:

\begin{eqnarray}
(\dot{G_{i}})_{out}\,=\,w_{i} \, \psi(t)
\end{eqnarray}
where $w_{i}$ is a 
free parameter that regulates the efficiency of the galactic
wind. The wind is assumed to be differential, i.e. some 
elements, in particular the products of 
SNe Ia, are lost from the galaxy more efficiently than others 
(Recchi, Matteucci $\&$ D'Ercole 2001; Recchi et al. 2002). 
This fact translates into slightly different values 
for the $w_i$ corresponding to different elements. Here we will
always refer to the maximum value of $w_i$. It should be pointed out
that the differential aspect of the wind has only a small influence on
the abundance ratio patterns (including the [$\alpha$/Fe], [Ba/Fe] and
[Eu/Fe] ratios). If, instead, a normal wind (one where all elements are lost
with the same efficiency) is used, the results will not change significantly.
The efficiency of the wind, on the other hand,  is crucial.
It is always high, but different for each dSph galaxy,
in order to account for the observational constraints. It is important
particularly to reproduce the [$\alpha$/Fe] ratio, the final total mass
and the present gas mass and to define the shape of the predicted
stellar metallicity distributions (for more details see LM03 
and LM04).

The initial mass function (IMF) is usually assumed to be 
constant in space and time in all the models and is 
expressed by the formula:

\begin{equation}
\phi(m) = \phi_{0} m^{-(1+x)}
\end{equation}

where $\phi_{0}$ is a normalization constant.
Following LM03 we assume a Salpeter-like IMF (1955)
($x=1.35$) in the mass range $0.1-100 M_{\odot}$.

In table 1 we summarize the adopted parameters for the models
of dSph galaxies.

\subsection{Nucleosynthesis prescriptions}

The nucleosynthesis prescriptions are essentially the same
as those adopted in LM04, with few modifications. Here we use,
as in LM04, Nomoto et al. 
(1997) for type Ia supernovae, but for massive stars
(M $> 10 M_{\odot}$) the yields of Woosley $\&$ Weaver (1995)
are used instead of those from Thielemann, Nomoto $\&$ Hashimoto
(1996) and Nomoto et al. (1997). This modification does not 
change the results obtained in the previous papers, and it is based
on the best fit to the data of very metal-poor
stars of our Galaxy found by Fran\c cois et al. (2004). The only 
difference is that the predicted [$\alpha$/Fe] ratios 
exhibit values somewhat higher, but still in agreement 
(actually a better one) with the data.

The type Ia SN progenitors are assumed to be white dwarfs in binary 
systems according to the formalism originally developed by Greggio
\& Renzini (1983a) and Matteucci \& Greggio (1986).

The major difference in the nucleosynthesis prescriptions 
is the inclusion of the yields of barium and 
europium in the code, following the procedure adopted 
by Cescutti et al. (2005), where details can be found.

{\bf Barium}

We assume two main different processes in the production of barium, taking
place in two different sites. The dominant s-process occurring in
low mass stars ($1 \le M/M_{\odot} \le 3$) and a low fraction of 
Ba being produced in massive stars through the r-process. The yields from
LMS are taken from Busso et al. (2001). The Ba yields resulting from
the r-process are assumed to be produced in massive stars in the mass
range 10 - 30 $M_{\odot}$. Travaglio et al. (1999) already predicted 
production of Ba in massive stars, but in the range 
8 - 10 $M_{\odot}$. Cescutti et al. (2005), on the other hand,
showed that the production of Ba in massive stars must be extended
to higher masses in order to fully reproduce the most recent 
observed trend of [Ba/Fe] in Galactic metal-poor stars (Fran\c cois et al.
2005). Following their procedure, we adopted the same yields 
for the r-processed Ba, which are shown in Table 2.

\begin{table} 
\begin{center}
\scriptsize  
\caption[]{The stellar yields of massive stars 
for barium and europium from Cescutti et al. (2005).}
\begin{tabular}{ccc}  
\hline\hline
\noalign{\smallskip} 
$M_{star}(M_{\odot}$) &$X_{Ba}$ &$X_{Eu}$\\   
\noalign{\smallskip}  
\hline
12. &$9.00*10^{-7}$ &$4.50*10^{-8}$\\
15. &$3.00*10^{-8}$ &$3.00*10^{-9}$\\
30. &$1.00*10^{-9}$ &$5.00*10^{-10}$\\
\hline\hline
\end{tabular}
\end{center}
\end{table} 
{\bf Europium}

The yields of europium are adopted assuming that its production occurs 
only through the r-process, which takes place in massive stars in a large
range of masses ($10 - 30 M_{\odot}$). Even though the production of
r-process elements is still matter of debate, in Cescutti et al. (2005) a
series of models adopting different nucleosynthesis prescriptions for Eu were
tested in order to reproduce the trend observed in Galactic metal 
poor stars. Here we adopted the yields, shown in Table 2, of the model which
best reproduced the data, i.e. their model 1.

\section{Results}

In order to follow the evolution of Ba and Eu in local dSph galaxies and,
in this way, to test the adopted nucleosynthesis prescriptions 
for these two elements we make use of the dSph chemical 
evolution models from LM03 and LM04 following the procedure 
and results described there.

In LM03 and LM04, we were able to fit the
observed [$\alpha$/Fe] ratios and the estimated final total mass and 
present day gas mass of six local dSph galaxies by varying the most 
important parameters, such as the SF efficiency and the 
galactic wind efficiency. The observational constraints were very 
well reproduced by models adopting very low SF efficiencies 
($\nu$ = 0.005 - 0.5 Gyr $^{-1}$) and high wind efficiencies 
(w$_i$ = 6-13). These two parameters together are the main responsibles
for the shape of the observed abundance ratios in dSphs. At low 
metallicities, at the beginning of the evolution of the system,
the [$\alpha$/Fe] ratios are relatively high ($\sim 
0.4$ dex) due to the almost instantaneous injection of $\alpha$-elements 
in the ISM
by massive stars which die in the form of SNe II. 
As the metallicity increases, the [$\alpha$/Fe] values start decreasing 
slowly, and soon after the first explosions of the SNe Ia they
go down fast to sub-solar values. This intense decrease after the first
SNe Ia explosions is caused by the injection of Fe in the ISM by these 
explosions and by the occurrence of the galactic wind triggered
by them. Since the galactic winds are very intense, with high 
efficiencies, they remove a large fraction of the gas reservoir
which feeds the SF and, consequently, the SFR drops to very low values.
With an almost negligible SF, the injection of $\alpha$- elements
in the ISM is almost halted, whereas
 the enrichment of Fe proceeds
for a very long time (a few Gyr) due to the long lifetime of
the stars responsible for its production and injection into the ISM.
Consequently, soon after the development of the wind, the 
predicted [$\alpha$/Fe] ratios suffer an abrupt decrease,
as it is observed. The final mass and present day gas mass
are also controlled by the SF and wind efficiencies: the higher the 
wind efficiency, the larger the mass of gas lost from the galaxy and the lower
the HI gas/total mass ratio. This scenario represents very well 
the six Local Group dSph galaxies analysed and is able to reproduce the
observational data. 

Here we adopt the same scenario as described above and the same range of
values for the main parameters of the models of LM04 (see 
table 1 for more details). By means of these models we follow the 
evolution of Ba and Eu 
and test the adopted nucleosynthesis prescriptions for these two 
elements. The predictions of the models are compared with [Ba/Fe], 
[Eu/Fe] and [Ba/Eu] as functions of [Fe/H].
 
\subsection{Europium}
\begin{figure}
\centering
\epsfig{file=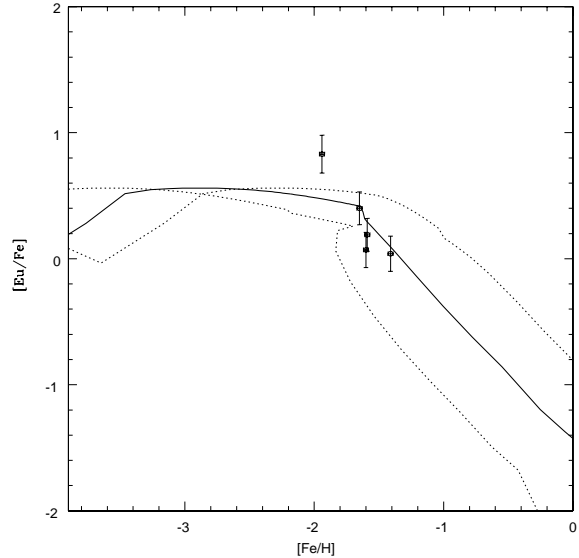,height=8cm,width=8cm}
\caption[]{[Eu/Fe] vs. [Fe/H] observed in Carina dSph 
galaxy compared to the predictions of the chemical evolution
model for Carina. The solid line represents the best model 
($\nu = 0.1\;Gyr^{-1}$, w$_i$ = 7) and the dotted
lines the lower ($\nu = 0.02\;Gyr^{-1}$) and upper 
($\nu = 0.4\;Gyr^{-1}$) limits for the SF efficiency.}
\end{figure}

The [Eu/Fe] ratio as a function of [Fe/H] observed in the four 
Local Group dSph galaxies is compared with the model predictions 
in the Figures 1 to 4 (Carina, Draco, Sculptor and Ursa Minor,
respectively). The predicted behaviour seen in the plots is the same for 
all galaxies: [Eu/Fe] is almost constant with supra-solar values
($\sim 0.5$ dex) until [Fe/H] $\sim -1.7$ dex (depending on the 
galaxy). Above this metallicity, the [Eu/Fe] values start decreasing
fast in Sculptor and Carina (there are no points at these metallicities
for Draco and Ursa Minor) similar to what is observed in the 
case of the [$\alpha$/Fe] ratio. This behaviour is consistent with the production of
Eu by r-process taking place in massive stars with $M > 10 M_{\odot}$.
Stars in this mass range have short lifetimes
and enrich
the ISM at early stages of galactic evolution
giving rise to high values of [Eu/Fe], since the production of Fe
in these stars is lower than in type Ia SNe occurring later. 
When the SNe Ia begin to occur,
the Fe abundance increases and, consequently, the [Eu/Fe] ratio
decreases, as one can see also in the data.
\begin{figure}
\centering
\epsfig{file=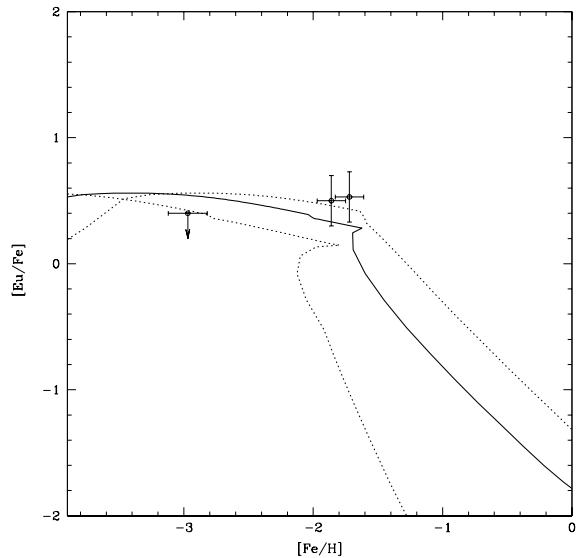,height=8cm,width=8cm}
\caption[]{[Eu/Fe] vs. [Fe/H] observed in Draco dSph 
galaxy compared to the predictions of the chemical evolution
model for Draco. The solid line represents the best model 
($\nu = 0.03\;Gyr^{-1}$, w$_i$ = 6) and the dotted
lines the lower ($\nu = 0.005\;Gyr^{-1}$) and upper 
($\nu = 0.1\;Gyr^{-1}$) limits for the SF efficiency.}
\end{figure}

\begin{figure}
\centering
\epsfig{file=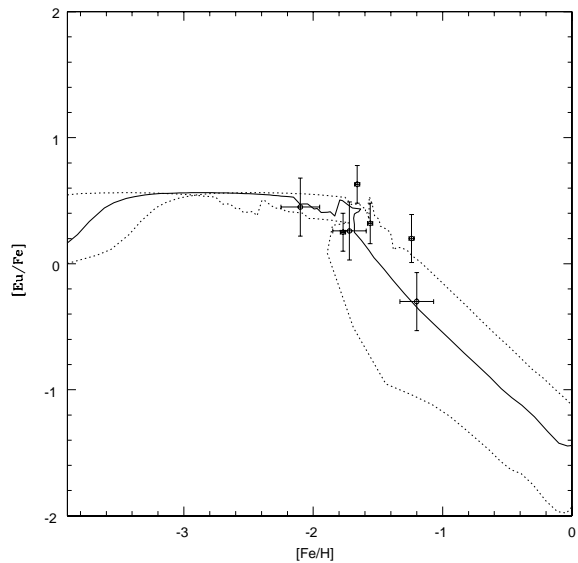,height=8cm,width=8cm}
\caption[]{[Eu/Fe] vs. [Fe/H] observed in Sculptor dSph 
galaxy compared to the predictions of the chemical evolution
model for Sculptor. The solid line represents the best model 
($\nu = 0.2\;Gyr^{-1}$, w$_i$ = 13) and the dotted
lines the lower ($\nu = 0.05\;Gyr^{-1}$) and upper 
($\nu = 0.5\;Gyr^{-1}$) limits for the SF efficiency.}
\end{figure}
The predicted [Eu/Fe] ratios in all four dSph galaxies  well reproduce 
the observed trend: an almost constant value 
at low metallicities, and an abrupt decrease starting at [Fe/H] 
$>$ -1.7 dex. In the model this decrease is caused not only by
the nucleosynthesis prescriptions and stellar lifetimes, but also 
by the effect of a very intense galactic wind on the star 
formation rate and, consequently,
on the production of the elements involved. 
In fact, since the wind is very efficient, a large fraction of the gas reservoir
is swept from the galaxy. At this point, the SF is almost halted
and the production of Eu goes down to negligible values. The injection
of Fe in the ISM, on the other hand, continues due to the large
lifetimes of the stars responsible for its production. The main 
result is an abrupt decrease in the [Eu/Fe] ratios, larger than the
one that one would expect only from the nucleosynthetic point of view if there
was no such intense wind. The abrupt decrease follows the 
trend of the data very well, especially in the case of Sculptor and
Carina. For these two galaxies there are stars observed with
metallicities higher than the one corresponding to the time when the wind
develops ([Fe/H]  $>$ -1.7 dex), and which are characterized by lower 
values of
[Eu/Fe], in agreement with our predictions. The observed
stars of the other two dSph galaxies, Draco and Ursa Minor, exhibit
[Fe/H] values which place them before the occurrence of SNe Ia, 
so it is not possible to verify if the abrupt decrease in the 
[Eu/Fe] occurs also in these objects. Only observations of more
stars will confirm the trend. It should be said again that
the same phenomena explain very well the [$\alpha$/Fe] ratios 
and the final total mass and present day gas mass observed in these 
galaxies (LM03, LM04). 
\begin{figure}
\centering
\epsfig{file=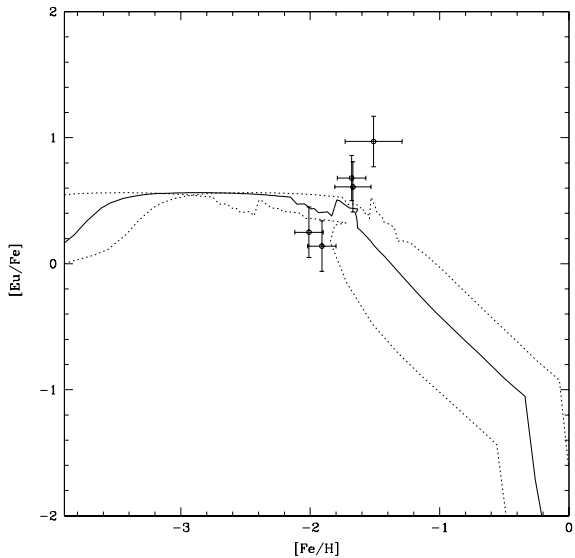,height=8cm,width=8cm}
\caption[]{[Eu/Fe] vs. [Fe/H] observed in Ursa Minor dSph 
galaxy compared to the predictions of the chemical evolution
model for Ursa Minor. The solid line represents the best model 
($\nu = 0.2\;Gyr^{-1}$, w$_i$ = 10) and the dotted
lines the lower ($\nu = 0.05\;Gyr^{-1}$) and upper 
($\nu = 0.5\;Gyr^{-1}$) limits for the SF efficiency.}
\end{figure}

The small differences in the SF and wind efficiencies do not affect 
strongly the
predictions of the models. As one can see in Table 1, the range of 
values for the SF efficiency is practically the same for Carina, Sculptor
and Ursa Minor ($\nu$ = 0.02-0.4, 0.05-0.5, 0.05-0.5 $Gyr^{-1}$, 
respectively), whereas Draco observational constraints are reproduced by
a model with lower values of $\nu$, $\nu$ = 0.005-0.1 $Gyr^{-1}$.
These values reflect in very similar curves for the first three  
galaxies and a curve for Draco with only a small difference, 
namely a [Eu/Fe] ratio which starts decreasing very slowly at metallicities
lower ([Fe/H] $\sim$ -2.0 dex) than in the other 
three galaxies. However, the abrupt decrease starts at a similar
point. The same similarity can be seen in the values of the wind efficiency:
Carina - w$_i$ = 7-11, Draco - w$_i$ = 6-10, Sculptor - w$_i$ = 11-15 -
and Ursa Minor - w$_i$ = 8-12. Only Sculptor is characterized by
a wind efficiency a little bit higher, but this fact does not influence
the pattern of the abundances significantly. They all exhibit an
intense decrease in the [Eu/Fe] ratio after the wind develops. The small
differences in the ranges of values for w$_i$ are related more directly
to the gas mass and total mass observed.

What should be highlighted is that the nucleosynthesis prescriptions
adopted here allow the models to reproduce very well the data, 
supporting the 
assumption that Eu, also in dSph galaxies, is a pure r-process 
element synthesized in massive stars in the range $M = 10-30 
M_{\odot}$, as it is in the Milky Way (see Cescutti et al. 2005
for a more detailed discussion). Besides that, the low SF efficiencies
and the high wind efficiencies are required also to explain
the [Eu/Fe] observed pattern, especially the abrupt decrease of 
the data in some dSph galaxies.

\subsection{Barium}

\begin{figure}
\centering
\epsfig{file=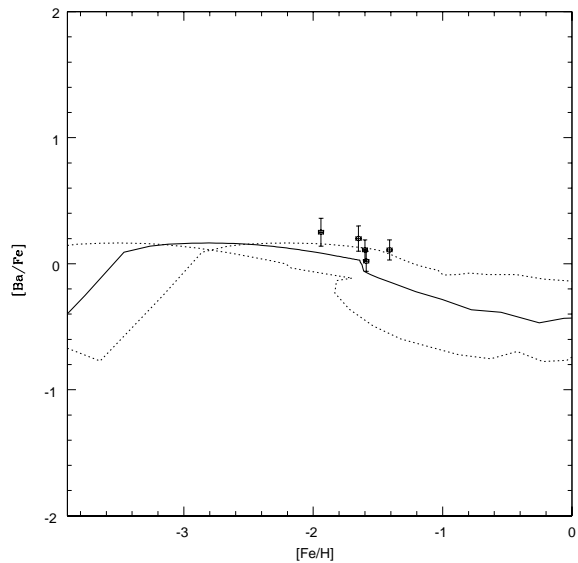,height=8cm,width=8cm}
\caption[]{[Ba/Fe] vs. [Fe/H] observed in Carina dSph 
galaxy compared to the predictions of the chemical evolution
model for Carina. The solid line represents the best model 
($\nu = 0.1\;Gyr^{-1}$, w$_i$ = 7) and the dotted
lines the lower ($\nu = 0.02\;Gyr^{-1}$) and upper 
($\nu = 0.4\;Gyr^{-1}$) limits for the SF efficiency.}
\end{figure}
The evolution of [Ba/Fe] as a function of [Fe/H] predicted by the models
and compared to the observed data in four 
Local Group dSph galaxies is shown in the Figures 5 to 8
(Carina, Draco, Sculptor and Ursa Minor,
respectively). One can easily notice that the predicted curves
exhibit a similar behaviour in all four galaxies: the predicted [Ba/Fe]
ratio increases fast at very low metallicities ([Fe/H] $<$ -3.5 dex),
then remains almost constant, close to the solar value, at low-intermediate 
metallicities (-3.5$<$ [Fe/H] $<$ -1.7 dex) and then starts decreasing soon 
after the occurrence of the galactic wind at relatively high
metallicities ([Fe/H] $>$ -1.7 dex). 
In this case, the decrease is not so intense as it is in the 
case of [Eu/Fe], due to the differences in the nucleosynthesis 
of Ba and Eu. 

\begin{figure}
\centering
\epsfig{file=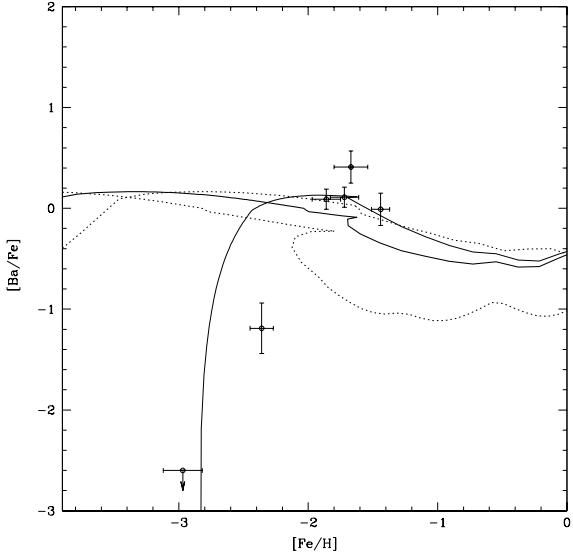,height=8cm,width=8cm}
\caption[]{[Ba/Fe] vs. [Fe/H] observed in Draco dSph 
galaxy compared to the predictions of the chemical evolution
model for Draco. The solid line represents the best model 
($\nu = 0.03\;Gyr^{-1}$, w$_i$ = 6) and the dotted
lines the lower ($\nu = 0.005\;Gyr^{-1}$) and upper 
($\nu = 0.1\;Gyr^{-1}$) limits for the SF efficiency. The thin line 
represents the best model without Ba production in massive stars.}
\end{figure}

The predicted shape of the [Ba/Fe] vs. [Fe/H] relation in dSphs 
can be associated 
to the two different Ba 
contributions, from stars in different mass ranges (high masses - 
10 to 30 $M_{\odot}$ - and low masses - 1 to 3 $M_{\odot}$).
In the low metallicity
portion of the plot 
the production of Ba is dominated by the r-process taking place
in massive stars which have lifetimes in the range from 6 to 25 Myr.
Therefore,  the [Ba/Fe] ratio increases fast 
reaching values above solar already at [Fe/H] $\sim$ -3.5 dex and 
stays almost constant up to [Fe/H]=-1.7 dex.
It is worth noting that the massive star 
contribution is more  clearly seen when the SF efficiency is low. 
In this regime, in fact, the stars are formed slowly and the 
difference between the contribution of stars of different 
masses is more evident, since the increase of the metallicity and 
the evolution of the galaxy proceed at a low speed. On the other hand,
when the SF 
efficiency is higher (like in the Milky Way), the early contribution of 
massive stars is 
more difficult to distinguish, because of the much
faster increase in metallicity. 

At low-intermediate metallicities (-3.5 $<$ [Fe/H] $<$ -1.7 dex),
the production of Ba is still the one by r-process taking place
in massive stars, in particular in those with masses around the lower 
limit for 
the r-process Ba producers ($\sim 10M_{\odot}$).

The contribution to s-process Ba enrichment from LMS 
(lifetimes from 3.8 x $10^{8}$ years 
to 10 Gyrs)  
affects significantly the predicted 
[Ba/Fe] ratio only after the onset of the wind, consequently 
only after the occurrence of the first SNe Ia. At this stage, the 
[Ba/Fe] starts to decrease rapidly, since the first SNe Ia are
injecting large amounts of Fe into the ISM. Together with the
enrichment of Fe, the SNe Ia release also large quantities of energy
in the ISM which gives rise to a galactic wind. As the galactic
wind starts, the SFR goes down to very low values and the production
of Ba is limited only to the LMS, especially those at the low mass
end. The injection of Ba in the ISM at this stage is, however,
not so effective due to the galactic wind which removes
a large fraction of the material freshly released in the hot 
medium (Ferrara $\&$ Tolstoy 1999, Recchi et al. 2001, 2004). 
The effect of the Ba production in LMS is particularly 
important to slow down the abrupt decrease
in [Ba/Fe] after the occurrence of the galactic wind. If this 
production is not taken in account, the [Ba/Fe] values after the
onset of the wind would go down faster to very low values. 

\begin{figure}
\centering
\epsfig{file=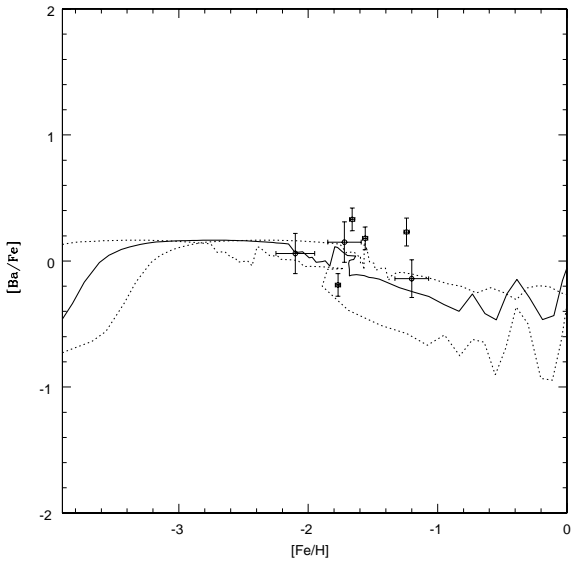,height=8cm,width=8cm}
\caption[]{[Ba/Fe] vs. [Fe/H] observed in Sculptor dSph 
galaxy compared to the predictions of the chemical evolution
model for Sculptor. The solid line represents the best model 
($\nu = 0.2\;Gyr^{-1}$, w$_i$ = 13) and the dotted
lines the lower ($\nu = 0.05\;Gyr^{-1}$) and upper 
($\nu = 0.5\;Gyr^{-1}$) limits for the SF efficiency.}
\end{figure}
One can see in the Figures 5 to 8 that the observational 
trends at high metallicities are very well reproduced by the model 
predictions 
supporting the assumptions made regarding the 
nucleosynthesis of Ba. As already 
mentioned the contribution from LMS to the enrichment
of Ba becomes important starting from  intermediate to high 
metallicities ([Fe/H] $>$ -1.9 dex), depending on the SF efficiency
adopted. In this 
metallicity range, the data of all four galaxies are very well 
reproduced, including the stars with low values of [Ba/Fe]
which should have formed soon after the onset of the 
galactic wind.
 
On the other hand, at low metallicities ([Fe/H] $<$ -2.4 dex)
only the observational trend of Carina and Sculptor are well 
fitted by the model predictions. In Ursa Minor and Draco there 
are a few stars
which exhibits a very low [Ba/Fe] 
($\sim$ -1.2 dex) at low [Fe/H] (Figures 6 and 8). These
points are well below the predicted curves and close to the values
of the Milky Way stars at similar metallicities, which are 
reproduced by a chemical evolution model with the same 
nucleosynthesis prescriptions adopted here (Cescutti et al. 2005) but with
a higher SF efficiency. In general, it seems like if the data for the solar 
neighbourhood show values of [Ba/Fe] 
lower than in the dSphs at the same metallicity, although this fact should 
be confirmed by more data. For the $\alpha$-elements is the opposite, dSph
stars show lower [$\alpha$/Fe] ratios than Galactic stars at the same 
metallicity (Shetrone \& al. 2001; Tolstoy et al. 2003).
LM03 and LM04 suggested that the difference in the behaviour of 
$\alpha$-elements in the Milky Way and dSphs should be ascribed to their 
different SF histories. In particular, the lower [$\alpha$/Fe] 
ratios in dSphs
are due to their low star formation efficiency which produces a slow 
increase of the [Fe/H] with the consequence of having the Fe restored 
by type Ia SNe, and
therefore a decrease of the [$\alpha$/Fe] ratios, 
at lower [Fe/H] values than in the Milky Way. This effect has been 
described in Matteucci (2001) and is a consequence of the time-delay 
model applied to systems with different star formation histories.

Therefore, in the light of what is said above, 
can we explain also the differences between 
the predicted [Ba/Fe] in 
dSph galaxies and in the Milky Way? Again, the SF efficiency is the major 
responsible parameter for this difference. In the Milky Way 
model the SF efficiency is much larger (10 - 100 times) than 
the ones adopted for the dSphs of the sample analysed here.
In the low efficiency regime, the contribution from 
LMS appears at lower metallicities than in the high SF regime, exactly 
for the same reason discussed for the [$\alpha$/Fe] ratios. As a 
consequence, we predict a longer plateau for the [Ba/Fe] ratio in dSphs 
than in the solar neighbourhood and starting at lower metallicities.
This prediction should in the future be confirmed or rejected by more data 
at low metallicities in dSph galaxies.

\begin{figure}
\centering
\epsfig{file=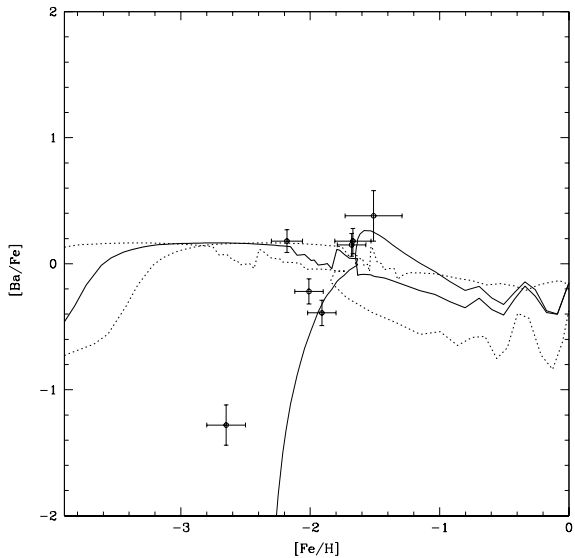,height=8cm,width=8cm}
\caption[]{[Ba/Fe] vs. [Fe/H] observed in Ursa Minor dSph 
galaxy compared to the predictions of the chemical evolution
model for Ursa Minor. The solid line represents the best model 
($\nu = 0.2\;Gyr^{-1}$, w$_i$ = 10) and the dotted
lines the lower ($\nu = 0.05\;Gyr^{-1}$) and upper 
($\nu = 0.5\;Gyr^{-1}$) limits for the SF efficiency. The thin line 
represents the best model without Ba production in massive stars.}
\end{figure}

\begin{figure}
\centering
\epsfig{file=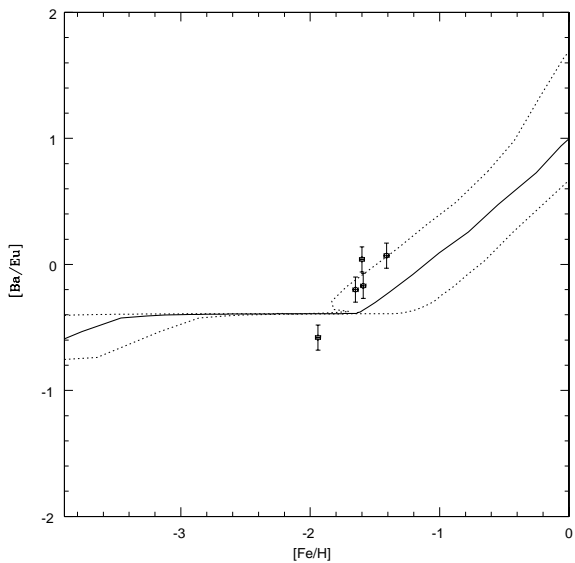,height=8cm,width=8cm}
\caption[]{[Ba/Eu] vs. [Fe/H] observed in Carina dSph 
galaxy compared to the predictions of the chemical evolution
model for Carina. The solid line represents the best model 
($\nu = 0.1\;Gyr^{-1}$, w$_i$ = 7) and the dotted
lines the lower ($\nu = 0.02\;Gyr^{-1}$) and upper 
($\nu = 0.4\;Gyr^{-1}$) limits for the SF efficiency.}
\end{figure}

Since there are no observed stars at  
low [Fe/H] in Carina and Sculptor while there are three stars 
(one with an upper limit) with very 
low [Ba/Fe] in Draco and Ursa Minor, one could argue that the Ba production 
from massive stars is not necessary.
To better see the effect of the r-process Ba production from massive stars,
we computed models suppressing this contribution. 
In such a case, the 
predictions of the models lie below all the observed data and are not 
capable of fitting the stars with low [Ba/Fe]. If, on the other hand, 
besides suppressing the contribution from r-processed Ba synthesised in 
massive stars one expands the production of s-processed Ba to stars 
with $M = 3 - 4 M_{\odot}$, with an yield $X_{Ba}$ = 0.5 x $10^{-6}$ 
in this mass range,
the models predict a trend similar to that observed 
(see the thin lines in Figures 6 and
8). The assumption  of Ba production 
by s-process in stars with $M = 3 - 4 M_{\odot}$ is justified 
by the fact that this production is predicted by models of stellar 
evolution (Gallino et al. 1998, Busso et al. 2001), even
though there are no such yields available in the literature. 
Besides that, Travaglio et al. (1999) 
suggested that the dominat production of Ba comes from stars with  
$2 - 4 M_{\odot}$.

As one can see from the thin lines in Figures 6 and 8, 
the increase of the [Ba/Fe] ratio occurs at
metallicities similar to those of  
the stars with low [Ba/Fe]. Besides that, this
model also reproduces the high values of [Ba/Fe] at high
metallicites and the [Ba/Eu] observed (see next section). In that sense, 
the observed low values of [Ba/Fe], if confirmed by more observations,
could be explained by a model with Ba produced only by s-process in 
stars with masses in the range  $M = 1 - 4 M_{\odot}$. The problem
is that a model with such yields overpredicts the
[Ba/Fe] at high metallicities in our Galaxy. Consequently, either
the production of Ba is not the same in the dSph Galaxies and in 
the Milky Way, as already suggested by Venn et al. (2004) or 
these stars are characterised by anomalous metallicities, as others 
observed in dSph galaxies (see Shetrone et al. 2001; Geisler et al. 2005).
Shetrone (2004) analysed the metal poor star Draco 119 (Fulbright, Rich 
$\&$ Castro 2004) which exhibits an upper limit for [Ba/Fe] $\sim$
-2.6 (an arrow in the Figures 2, 6 and 10) and suggested that this 
specific star might be contaminated by 
inhomogeneus mixing, since other stars observed in Draco at similar 
metallicities present higher values for [Ba/Fe] ($\sim$ -1.0 dex). 
These other stars, 
however, seem to indicate that these low values of [Ba/Fe] 
are  not uncommon in Draco and, as a possible consequence, that the 
production of Ba in Draco (maybe also in other dSph galaxies) is
not the same as it is in our Galaxy.

Only more
observations of stars with similar metallicities (lower than 
[Fe/H] $\sim$ -2.0 dex) in dSph galaxies could solve this problem.

\begin{figure}
\centering
\epsfig{file=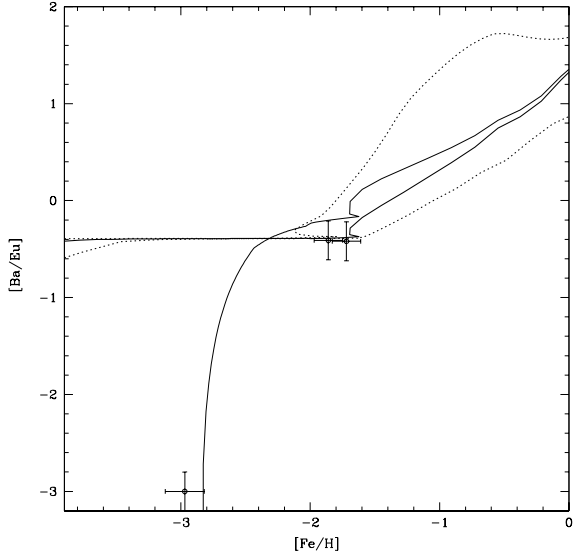,height=8cm,width=8cm}
\caption[]{[Ba/Eu] vs. [Fe/H] observed in Draco dSph 
galaxy compared to the predictions of the chemical evolution
model for Draco. The solid line represents the best model 
($\nu = 0.03\;Gyr^{-1}$, w$_i$ = 6) and the dotted
lines the lower ($\nu = 0.005\;Gyr^{-1}$) and upper 
($\nu = 0.1\;Gyr^{-1}$) limits for the SF efficiency. The thin line 
represents the best model without Ba production in massive stars.}
\end{figure}


\subsection{The ratio [Ba/Eu]} 

The comparison between the observed [Ba/Eu] as a function of [Fe/H] 
and the predicted curves for the four dSph galaxies is shown
in Figures 9 to 12. The models predict a similar pattern for all four
galaxies: an almost constant sub-solar value at low metallicities
([Fe/H] $<$ -1.7 dex) and, after that, a strong increase. This 
pattern is explained again by the adopted nucleosynthesis
and by the effect of the galactic wind on the SFR and, consequently,
on the production of Ba and Eu. At the early stages of evolution, the
high mass stars provide the major contribution to the enrichment of the
ISM medium. Since Ba and Eu are both produced by the r-process taking place
in massive stars,
they both are injected in the ISM when the gas metallicity is still low. 
The difference
is that Eu is considered to be a pure r-process element, while the
fraction of Ba that is produced by the r-process is low and its 
bulk originates instead from LMS.
This fact translates into the sub-solar pattern observed in the predicted
curves: more Eu than Ba is injected in the ISM at low metallicities, at an
almost constant rate. When the LMS start to die and the first SNe Ia 
start exploding the scenario changes significantly. The LMS
inject a considerable amount of Ba into the ISM causing an increase in the
[Ba/Eu] ratio. Besides that, the energy
released by the SNe Ia contributes to the onset of the galactic wind. 
Since the wind is very intense, it removes from the galaxy a large 
fraction of the gas reservoir which feeds the SF. Consequently, 
the SFR drops down considerably and also the  production of Eu 
by massive stars, because the number of new formed stars is almost 
negligible. Barium, on the other hand,
continues to be produced and injected in the ISM by the LMS
(s-process). This fact induces the increase of [Ba/Eu] to be even more
intense, as one can see in the predicted curves (Figures 9 to 12).

\begin{figure}
\centering
\epsfig{file=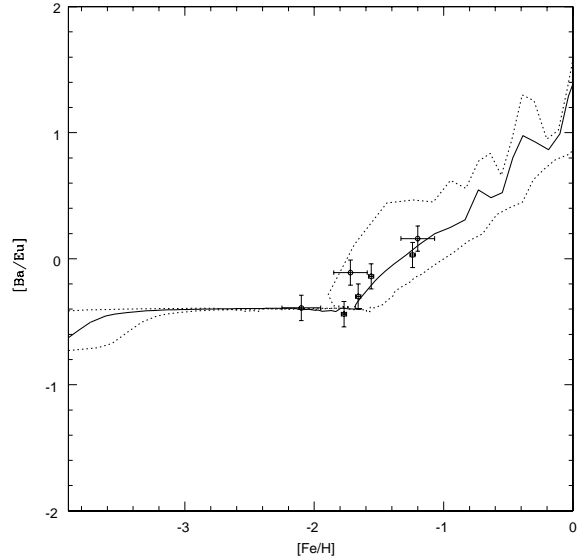,height=8cm,width=8cm}
\caption[]{[Ba/Eu] vs. [Fe/H] observed in Sculptor dSph 
galaxy compared to the predictions of the chemical evolution
model for Sculptor. The solid line represents the best model 
($\nu = 0.2\;Gyr^{-1}$, w$_i$ = 13) and the dotted
lines the lower ($\nu = 0.05\;Gyr^{-1}$) and upper 
($\nu = 0.5\;Gyr^{-1}$) limits for the SF efficiency.}
\end{figure}
The observed trend is very well reproduced by the predicted curves
in all four galaxies, especially in the case of Carina and Sculptor
(Figures 9 and 11, respectively).
The abundance pattern of these two galaxies not only exhibits the 
"plateau" at low metallicities, but also the sudden observed increase of
[Ba/Eu] after the onset of the wind, suggesting that the adopted
nucleosynthesis prescriptions for both Ba and Eu are appropriate
and that the scenario described by the chemical evolution models
is suitable to explain the evolution of these galaxies. In the case
of Draco and Ursa Minor (Figures 10 and 12, respectively), there are
no stars with metallicities larger than [Fe/H] $\sim$ -1.7 dex, the one
characteristic for the onset of the galactic wind. Therefore, one 
cannot verify if this scenario (after the occurrence of the wind)
holds also for these systems. However, the "plateau" is very well 
reproduced, even though there is some dispersion in the data, 
especially in the case of Ursa Minor.

\begin{figure}
\centering
\epsfig{file=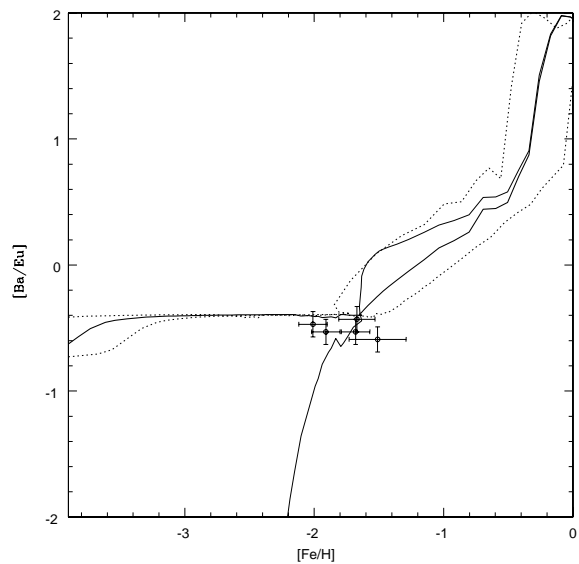,height=8cm,width=8cm}
\caption[]{[Ba/Eu] vs. [Fe/H] observed in Ursa Minor dSph 
galaxy compared to the predictions of the chemical evolution
model for Ursa Minor. The solid line represents the best model 
($\nu = 0.2\;Gyr^{-1}$, w$_i$ = 10) and the dotted
lines the lower ($\nu = 0.05\;Gyr^{-1}$) and upper 
($\nu = 0.5\;Gyr^{-1}$) limits for the SF efficiency. The thin line 
represents the best model without Ba production in massive stars.}
\end{figure}
It is important to stress that the predicted [Ba/Eu] reproduces
all the observed trends, and that no star, given the uncertainties
(with the exception of Draco 
119, which exhibits a very uncertain value due to the limits on 
Ba and Eu abundances),
lies outside the predictions, as it was the case for the two stars 
with very low [Ba/Fe]. This fact suggests strongly 
that the outsider stars must be examined separately.

\subsection{The Sagittarius dSph galaxy}  

In this section, we present the predictions for [Ba/Fe], [Eu/Fe] and 
[Ba/Eu] as functions of [Fe/H] in Sagittarius dSph galaxy. Even
though there are only two stars (Bonifacio et al. 2000)
observed with Ba and Eu, it is 
interesting to compare the predictions of the models to the data and to
analyse how these ratios would behave in this dSph galaxy. As
mentioned in LM04, the Sagittarius dSph galaxy exhibits chemical
properties which distinguish this galaxy from the other Local Group dSph 
galaxies. In particular, the SF efficiency (required to reproduce the
observed [$\alpha$/Fe] ratios) and the predicted metallicity 
distribution of this galaxy differ a lot from the other dSph 
galaxies analysed - Draco, Carina, Ursa Minor, Sextan and Sculptor - 
being more similar to the values assumed for the Milky Way disc. The 
required SF 
efficiency is much higher ($\nu$ = 1 - 5 Gyr$^{-1}$
compared to $\nu$ = 0.01 - 0.5 Gyr$^{-1}$) and the stellar metallicity
distribution exhibits a peak at higher metallicities ([Fe/H] $\sim$ -0.6 
dex) than the other dSph galaxies ([Fe/H] $\sim$ -1.6 dex) and 
close to the one from the solar neighborhood.  As a consequence, one 
would expect also [Ba/Fe], [Eu/Fe] in Sagittarius to be different 
from the patterns observed in the other four dSph and more similar to 
those observed in the metal-poor stars of the
Milky Way. 

In order to predict the evolution of Ba and Eu as functions of Fe, we 
made use of the Sagittarius dSph model as described in LM04, without 
any changes in the most important parameters, such as SF efficiency 
and wind efficiency, and with the same nucleosynthesis prescriptions 
adopted for the other dSphs. This procedure is justified 
by the fact that no modifications were required for the LM04 models 
of the other galaxies (Carina, Draco, Sculptor and Ursa Minor) to
fit the observed [Ba/Fe], [Eu/Fe] and [Ba/Eu].

\begin{figure}
\centering
\epsfig{file=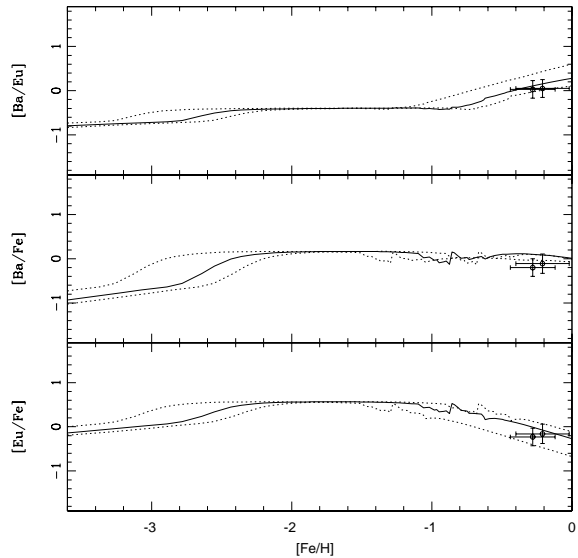,height=8cm,width=8cm}
\caption[]{The predicted evolution of Ba and Eu as function of [Fe/H] for
Sagittarius dSph galaxy compared with the data. The solid line represents
the best model  ($\nu =
3\;Gyr^{-1}$, w$_i$ = 9) and the dotted lines the lower ($\nu =
1\;Gyr^{-1}$) and upper  ($\nu = 5\;Gyr^{-1}$) limits for the SF
efficiency. } 
\end{figure}

In Figure 13, the predictions for Sagittarius dSph galaxy model for 
[Ba/Fe], [Eu/Fe] and [Ba/Eu] are shown in comparison with the data. 
As one can clearly see, all three predicted ratios reproduce very well the 
data and exhibit significant differences (in particular [Ba/Fe])
when compared to the predictions (and observations) for 
the other dSph galaxies. The decrease in the [Eu/Fe] at relatively 
high metallicities ([Fe/H] $\sim$ -1.7 dex) observed in the four 
dSph galaxies, and attributed to the effect of the galactic wind on 
the SFR, is less intense in the case of Sagittarius. Moreover, one cannot 
see the high values of [Ba/Fe] at low metallicities ([Fe/H] 
$<$ -3.0 dex ), which were explained as an effect of the low SF 
efficiency.  Also, the predicted [Ba/Eu] ratios do not show the 
almost constant  "plateau" observed at low and intermediate metallicities 
in the other dSph  galaxies. All these differences can be also 
found when one compares the pattern of these ratios in dSph galaxies 
with those in the metal-poor stars of the Milky Way. The differences 
between the predictions of Sagittarius and the other dSph and
the similarities with the Milky Way can be attributed to the high
values of the SF efficiency adopted for Sagittarius when compared 
to the other dSph galaxies. These high values, in fact, are more 
similar to the values generaly adopted for the solar neighborhoud 
(Chiappini, Matteucci $\&$ Gratton, 1997). 
Consequently, one could suggest that the chemical evolution of 
Sagittarius follows roughly that of the Milky Way disc at the solar 
neighborhoud in contrast to the other dSph Galaxies,
which exhibit a much slower chemical evolution.

\section{Summary}

By means of a chemical evolution model which is able to reproduce
several observational constraints of the Local Group dSph galaxies 
(such as [$\alpha$/Fe], present day gas mass, estimated final total 
mass, metallicity distribution) we followed the evolution of Ba and 
Eu as a function of Fe in
these galaxies in order to verify some assumptions regarding
the production of these elements. The model galaxies are 
specified by the SF prescriptions, such as the number, epoch
and duration of the episodes, the SF efficiency and also by the
wind efficiency. These two last parameters are the main responsible
in defining the shape of the abundance ratio patterns and for the
depletion of the gas content of the galaxy. The SF efficiency must 
be characterised by low values ($\nu$ = 0.005 - 0.5 Gyr$^{-1}$) 
in order to reproduce the observed data whereas the efficiency 
of the galactic winds must be high ($w_i$ = 6 - 13). The effects
of a low SF efficiency and an intense wind efficiency on the production
of Ba and Eu coupled to the nucleosynthesis prescriptions suggested 
by Cescutti et al. (2005) and adopted in this work enable us to
reproduce very well the observed trends of [Ba/Fe], [Eu/Fe] and
[Ba/Eu] as function of [Fe/H]. This agreement also allowed us 
to suggest some constraints on the formation and production 
of Ba and Eu.

The main conclusions can be summarized as follows: 
\begin{itemize}

\item
the observed [Eu/Fe] ratio is very well reproduced by the models for all
four dSph galaxies with the assumption that this element is synthesized
by r-process in massive stars in a defined range of masses 
(10-30 $M_{\odot}$). The pattern of [Eu/Fe] is explained by the 
different sites of production of these
elements, Fe being mainly produced in SNe Ia on long time-scales and Eu in 
SNe II on short time-scales,  and by the effect of the galactic winds
on the SFR. At the early stages of evolution the major contributor
to the ISM enrichment are the massive stars,  
giving rise to a high [Eu/Fe] at low [Fe/H] ($<$ -2.0 dex). 
When the first SNe Ia start exploding and restoring the bulk of Fe
then the [Eu/Fe] ratio starts decreasing (time-delay 
model, Tinsley 1980; Greggio $\&$ Renzini 1983b; Matteucci $\&$ 
Greggio 1986; Matteucci 1996). With the energy released by 
these explosions a 
galactic wind is triggered and, since it removes a large fraction 
of the gas reservoir which fuels the
SF, the SFR drops down considerably. As a consequence, the enrichment
of Eu is almost halted but the one of Fe continues for a long
time giving rise to very low values of [Eu/Fe], as observed;

\item
the predictions of the models for [Ba/Fe] reproduce very well the 
observed data in all four dSph galaxies with the exception of two
stars (one in Draco and another in Ursa Minor) which exhibit very low
values of [Ba/Fe] at low metallicities ([Fe/H] $<$ -2.4 dex) and could 
be anomalous stars. The 
[Ba/Fe] pattern is explained
by the different contributions to the production of Ba and by the 
the effect of the galactic winds on the SFR. Ba is assumed to be produced
mainly by s-process in LMS (1-3 $M_{\odot}$), but also by r-process in 
massive
stars in the range 10 - 30 $M_{\odot}$. This last production is 
more important at early stages of galactic evolution, at low 
and intermediate metallicities, and is responsible
for the high values observed in dSphs at low [Fe/H] and to maintain the 
predicted [Ba/Fe] almost constant at supra-solar values at 
intermediate metallicities. The s-processed Ba produced in 
LMS becomes important later, at higher metallicities 
([Fe/H] $>$ -1.9 dex), when the
wind develops. 

\item
the low values of [Ba/Fe] at low [Fe/H] observed in Draco 
and Ursa Minor, 
if confirmed by more 
observations, could be explained by a model in which Ba is produced
only by s-process occuring in stars with masses in the range 
1 - 4 $M_{\odot}$. 
However, this production of Ba is different from
the one adopted by a Milky Way model which successfully reproduces the 
observations (Cescutti et al. 2005). 
Consequently, if the low values of [Ba/Fe] in dSph galaxies
are real, then the production of Ba could be not the same in the dSph 
galaxies 
and in the Milky Way. On the other hand, if no other star is observed
with such low values, their abundance pattern could be anomalous.
This problem could be solved only with  more observations of stars 
in the same metallicity range of the star with reported low [Ba/Fe] 
ratio; 

\item
the observed [Ba/Eu] as a function of [Fe/H] is very well reproduced
by the models with the adopted assumptions regarding the 
nucleosynthesis of Ba and Eu. The different sites for the production of
these elements and the effects of the galactic winds on the SFR are 
again the main responsibles for the pattern observed in [Ba/Eu]. The 
sub-solar "plateau"
at low metallicities is caused by the injection into the ISM of Ba and 
Eu by massive stars, in different fractions: Eu is assumed to be a pure 
r-process element, whereas the fraction of r-processed Ba is low. This 
gives rise to a sub-solar [Ba/Eu]. When the first SNe Ia explode and the
wind develops, there is an abrupt change and the [Ba/Eu] suffers an
intense increase, due to the injection in the ISM of Ba by LMS 
and to the decrease in the SFR. With a very low SFR, the production of
Eu is almost halted, but the one of Ba continues since the LMS
have a long lifetime (from several $10^{8}$ years to several Gyrs). 
Therefore, [Ba/Fe] increases considerably, as observed in the dSph stars;

\item
the nucleosynthesis prescriptions adopted in this work are the same
adopted in a chemical evolution model for our Galaxy which reproduces 
very well the [Ba/Fe], [Eu/Fe] and [Ba/Eu] trends observed
in Milky Way (Cescutti et al. 2005). This agreement, 
coupled to the one achieved here (with red giant stars observed 
in local dSph galaxies), strongly suggests that the assumptions 
regarding the formation and production of Ba and Eu
are quite reasonable;

\item
we also compared the data of two observed stars with the predicted 
evolution of Ba and Eu as functions of Fe in
Sagittarius using the same model as described in LM04 and with the same
nucleosynthesis prescriptions adopted in this work. The predictions 
exhibit significant differences when compared to the predictions and 
observations of the other four dSph galaxies, but similarities to the 
metal-poor stars of the
Milky Way. Both facts can be attributed to the much higher values of SF
efficiencies adopted for the Sagittarius galaxy when compared to the other
galaxies. This galaxy is much larger and more massive than the
other dSphs and could be characterized by a chemical evolution more 
similar to the one of the solar neighborhoud in the Milky Way disc 
than to the one of the other dSph galaxies;

\item
finally, we are able to explain the different behaviour of the [Ba/Fe] 
and [Eu/Fe] in the dSph galaxies and in the Milky Way (here one should 
place Sagittarius together with the Milky Way instead of with the 
dSph galaxies). The higher predicted and  observed values 
(although these should be confirmed by more data) of these ratios at low 
metallicities in dSph galaxies are due to the much less efficient 
star formation adopted for these galaxies. In this star formation 
regime, in fact, the metallicity increases more slowly and the 
different contributions for the Ba enrichment of the ISM appear
at lower metallicities than in the Milky Way. 
For the same reason the [$\alpha$/Fe] ratios in dSphs are lower than the 
same ratios in the Milky Way at the same metallicities, as suggested 
already by LM03 and LM04. 

\end{itemize}

\section*{Acknowledgments}
G.A.L. acknowledges financial support from the Brazilian agency 
FAPESP (proj. 04/07282-2). F.M. acknowledges financial support  
from INAF Project ``Blue Compact Galaxies: primordial helium and chemical 
evolution'' and from COFIN2003 from the Italian Ministry for Scientific 
Reasearch (MIUR) project ``Chemical Evolution of Galaxies: interpretation 
of abundances in galaxies and in high-redshift objects''.

\bsp

\end{document}